# Philosophy of Science and Educational Research: Strategies for Scientific Effectiveness and Improvement of the Education

Omar a. Ponce[a], José Gómez Galán[b] and Nellie Pagán-Maldonado[c]

[a] Professor (Metropolitan University, AGMUS, Puerto Rico-United States). um_oponce@suagm.edu
[b] Research Professor and Director of CICIDE (Metropolitan University, AGMUS, Puerto Rico-United States & Catholic University of Avila, Spain). jogomez@suagm.edu & jose.gomez@ucavila.es
[c] Professor (Metropolitan University, AGMUS, Puerto Rico-United States). npaganm@suagm.edu

**Abstract**

This article is a theoretical study on the effectiveness of educational research in the context of philosophy of science. This topic of discussion, in the area of educational research, has been the subject of intellectual debate and arises again at the beginning of the 21st century. This article outlines the challenges and opportunities for scientific effectiveness facing educational research if it aspires to contribute to the ideal of an education of excellence and quality. Nine strategies to improve scientific effectiveness in educational research are identified and discussed. As a conclusion, it is argued that the foundations of contemporary educational research need to be revisited and reformulated, parallel to the new concepts present in the philosophy of science, to face the new problems present in our society[*].

## 1. Introduction

Philosophical thinking centered on science has been an object of intense debate for centuries. Thus, when we speak today of philosophy of science, we refer fundamentally to any reflection produced around the scientific methodology and its results. It faces, from the point of view of reason, the nature of science and the philosophical problems generated around it, with a direct implication in its results and applications.

There have been multiple currents of philosophy of science encountered during the twentieth century. It is possible to find from; the logical positivism of the Vienna Circle, developed in parallel to the revolution in the field of theoretical physics with Einstein's studies and deeply disrupting traditional philosophical thinking centered on empiricism and inductivism [1] [2] [3], to Feyerabend's new proposals against the method, which speak of methodological and epistemological anarchism or the absence of a scientific method [4] [5], as well as Ernst Cassirer's classic contributions in the sphere of neo-Kantianism [6] [7], Ernst Mach and the positivism [8] [9] [10] or Popper and his critical rationalism [11] [12]

Of course, when we focus on the social sciences, and especially on the educational sciences, the problem becomes even more complicated. Since we are in a diffuse scientific field that escapes the fundamental presuppositions of the natural sciences, and whose study of the problems, with an indeterminate number of variables involved, becomes extremely complex. It would even be necessary to reflect in essence on whether we can understand education as a science [13]. However, regardless of their epistemological bases, the truth is that educational research is fundamental in the scope of knowledge necessary for the development of educational processes in all its dimensions. Education is one of the main pillars of our society [14].

Throughout its history, education has been an essential instrument for social and economic development in public policy in many countries. In order to respond to the ideals of social and economic development, education has had to think and rethink its managerial and organizational structures, programs and curricula, teaching practices, skills and recruitment requirements of educators, and their mechanisms to display

---

evidence of compliance. These changes are evident under the name of educational reforms. Education has not been a working reality or a stable or uniform field of scientific research. The needs and educational realities of industrial society seem similar, but they are not identical to the needs and educational realities of the global, technological and computer society in which we live. There are many interests, needs and protagonists that converge in this realm of reality that we call education. Generating educational research that contributes to the development of the profession has been, is, and will continue to be a challenge in the field of education. The philosophical study of educational research makes it clear that educational research is much more than research methods. The methods for conducting effective educational research are important, but equally so is having a defined north of educational research.

Thus, based on a critical and post-structuralist analysis, we present different strategies focused on scientific effectiveness that address the challenges we face today. We talk about topics that need to be studied and analyzed in order to improve the quality of education and, consequently, society. That would reflect a new way of understanding the philosophy of science applied to the field of education sciences.

## 2. Development: Strategies for Scientific Effectiveness in Educational Research

The following development strategies are opportunities to improve the practices that defined educational research in the 20th century [15]. These strategies certainly present great challenges, but also great opportunities for scientific effectiveness in educational research:

Strategy 1: Define and demarcate the concept of educational research
It is interesting that academics and educational researchers from different countries agree on the need to define and demarcate the concept of educational research by the impact on the scientific effectiveness that this entails [16] [17] [18] [19] [20] [21]. Educational research is too broad and ambiguous, which in practice needs to be defined and structured. If this is achieved, it is easier to define the roles, participation and scientific contributions of each of the models of educational research. For example, two purposes of research in many disciplines of study are description and causal relationship. Each of these categories makes it easier to identify the type of research that best applies, the expected output of that research, the quality criteria to evaluate it, and the application to the practice or development of public policy.

The opportunity of scientific effectiveness at this historical moment is to define the path that educational research must take so that each researcher can understand where to go [17] [20] [22]. This should contribute to scientific continuity, regardless of government changes that control education. The lack of a common vision on educational research affects its effectiveness because there are no common and articulated efforts of scientific research. This has raised questions about the quality, scope and usefulness of educational research. Educational research has no defined model or paradigm on how to address the complexity of education. Educational research needs to define, claim its particularities and its own identity as a field of scientific research [19] [20] [22] [23] [24] [25].

Strategy 2: Define the education construct as a research phenomenon
Some of the problems of effectiveness of educational research emanate from the indefinition of the construct education as a phenomenon of scientific research [16] [17] [20] [21] [25] [26]. For example, what is education or what is the best way to investigate it? Education is a phenomenon for which there is no universally accepted definition [16] [17] [20] [21] [25]. From its beginnings as a profession, education has been investigated from the premise that it is a natural phenomenon or that it is a social phenomenon. These visions of education emerge with the methods of quantitative and qualitative research that were adopted in the scientific investigation of education. The emphasis of educational research from these viewpoints was the method and not the construct of education. The result of this is the absence of a common view of education as a scientific research phenomenon. The problem of educational research in the 21st century is how the field of education is defined [16] [21] [25] [27]. The problems of effectiveness of educational research have to do with the nature of the data, their origin and how they emerge [28]. For this, educational research needs to empirically build the education construct [26] and develop a universal language to capture educational reality [19] [29]. The effect of the indefiniteness of the education construct is observed in the need to define the concept of educational research to increase its scientific effectiveness [16] [19] [28] [29] [30]. Touriñan [29] argues that scientifically investigating the education construct provides to understand the theory and practice of education and to generate the theory that explains education. In other words, research on the knowledge of education and research on education as a realm of reality. Educational research needs not only to describe, explain, understand, interpret and transform education from its methodologies, but also from its concepts and constructs.

Strategy 3: Conduct research that is useful for education

Another source of problems of effectiveness with educational research emanates from its relation with the practice of the profession [20] [25] [31]. For some, the relationship between research and practice has been controversial [32], imperfect and sometimes non-existent [25] [33] [34]. This has raised questions about the quality and usefulness of educational research. It is argued that there are too many studies that contribute little or nothing to the solution of the real problems of education [20] [25] [31] [35] [36]. In order for research to contribute to the improvement of education, it must focus on its problems and its real needs [20] [23] [24] [31] [36] [37] [38]. The call in the 21st century is to an education based on the results of educational research [34]. The usefulness of knowledge emerges as a relevant topic on the effectiveness of educational research [35].

The usefulness of knowledge is framed in the link that must have educational research with the practice of the profession. An effective educational research in the 21st century is considered to be one that generates theory, which guides the practice of the profession or which informs its educational policies [31] [33] [35] [39] [40]. This fact is clearly linked to the absence of a vision of what educational research means. The usefulness of knowledge is also linked to the scope of research. The scope of research means that there are too many small-scale studies. The reduced scope of the study does not facilitate responding to the real problems that are manifested in schools, school districts or education systems. For a study to contribute significantly to the improvement of education, researchers must address the phenomena in their full extent and dimension. The tendency to delimit studies under the argument of being able to investigate the research problem produces studies without transcendence [20] [25]. The usefulness of knowledge is considered an opportunity to link educational research with the needs of public education [33] [35] [41]. The interest of connecting educational research with the classroom and increasing the usefulness of knowledge is a subject that brings interesting reflections. Who determines the usefulness of knowledge, the researcher, educators, educational administrators or politicians on duty?

Strategy 4: Increase the validity of educational research

Educational research is largely ex post facto. Ex post facto research sometimes resorts to the lives and experiences of the protagonists of education to generate knowledge. This raises questions about the possibility of knowing reality and human behavior [42]. Educational research needs to eliminate from its studies the opinions to demonstrate the facts if it aspires to be scientific. It needs to replicate further, studies as a form of validation and confirmation of their knowledge [39]. Human complexity is of such magnitude that educational research has to take place in broader contexts where the real product of education can be appreciated [19]. This leads to two considerations: (a) Research in educational institutions. The value of educational and managerial practices is that they have to occur in educational settings where it is possible to observe, describe and measure the effect of these on education and institutional effectiveness. This research informs the creation of educational policies and the decision-making of all its constituents [43]. There is a need, and it is possible to conduct empirical studies to prescribe what education should be and understand its nature. Such studies should occur as close to the classroom [44]. (b) Research outside the context of educational institutions. There is a need to develop educational research away from the classroom where the products of education are actually observed. Educational research must move away from educational contexts where education takes place to understand other relationships about how education occurs [30].

Strategy 5: Increase the generalization of educational research

The issue of generalization of knowledge is recurrent in the allocation of funds for educational research. There are the following consensus on generalization in educational research: it cannot be the search for laws of universal application; does not imply that the knowledge of a particular situation can inform upon another similar situation and the deep description of the educational phenomenon under study is fundamental to be able to identify the similarities and differences that allow the knowledge of one situation can help to inform another [45]. In the field of education students are studying educational practices, policies and behaviors. So generalizing in this context entails the following considerations [18]: (a) Human behavior can be influenced by the biology of the person, but also by their culture. Education changes culture. So the search for these cultural patterns is necessary to be investigated and discovered. (b) Generalizing always implies some kind of inference when applying. There are numerous arguments about the investigation of people or individuals, but much of the phenomena of education occur and are investigated in groups. Generalization must be parallel between individuals and between groups as in the physical sciences. (c) The generalization of the findings can be increased by considering the following aspects of knowledge production and their accuracy:

definition and careful delineation of the phenomenon being investigated, determining the functions and structures of the causal relationships being studied, identifying the factors that are causal in the situations that are investigated, distinguish which factors are causal and which are coincidences, precise measurement, develop and validate theories and hypotheses, and connect causes with effects; (d) Defining the educational context and the phenomenon should help to better understand the duration of the data. In education, the important findings of a decade may lose their full relevance in the next decade and with the next generations of students because socio-economic, cultural, or political conditions change. In the physical world, many findings last for life. This facilitates the accumulation of knowledge. In education, social, cultural and political conditions can make data lose relevance [46].

Strategy 6: Careful selection of the research design to be used
The premises of science are two: (a) That reality is intangible. This does not mean that it is superficial. (b) Knowledge is not easy to achieve, but the human being must fight against superstition and simple solutions presented by the mind [47]. In contemporary culture of scientific research to generate knowledge, procedures are essential to differentiate between scientific knowledge and ordinary or daily knowledge of daily living. The result is the belief that the method controls research actions. The two limitations that arise when generating objective knowledge are the following: each conclusion is restricted to the method and the premises that generate it, and each research is developed from the perspective that the researcher selects to approach the phenomena they study and the interpretation they give to them [19]. In the 21st century, the selection of the research method must be aligned and respond to the scope it presents to capture the complexity of education and generate the data that is needed. The study of education does not have to be limited to experimental research as long as a solid case is developed or solved [22]. So many quantitative, qualitative and mixed methods present strengths and challenges when investigating the complexity of education. These strengths and challenges must be known in order to maximize their potential in the search for useful knowledge for education [20] [25].

Strategy 7: Explore new models for the study of causal relationships and educational effectiveness
The dominant stance in literature is that causal relationship research is an important issue in education because it impacts the effectiveness of schools. It is also understood that educational research has succeeded in advancing the study of causal relationships because it has refined its research methods. Most of the recommendations documented in this book are intended to strengthen methods and designs of educational research. The dominant view is that educational research must be pragmatic and respond to the needs of education. Technology in educational research is a tool to obtain more precise data and to study the causal relationships. In spite of these advances, the argument arises that the investigation of causal relations seems to be an ideal unattainable due to the complexity in education.

The critique of causal research in the field of education is that it does not capture the complexity of school reality. Much of the causal research is part of the results of standardized tests that seek to connect educational activities (i.e. teaching techniques, curricula, programs) with results [37]. The premise is that learning outcomes can be explained in causal, linear and logical relationships between activities and their products or outcomes. Causal research approaches the educational phenomenon by isolating the study variables and extracting them from the context and the educational reality. In doing so, educational processes are obviated, other important causal variables can be hidden, and school reality can be distorted [48] [49]. For example, explaining the achievement of educational standards excludes the issue of diversity that occurs in schools. It is for this reason that causal research alone cannot be considered a complete approach to the study of the effectiveness of schools.

The complexity of education is an issue that is discussed in relation to the effectiveness of educational research [20] [21] [25] [48] [49]. In order to advance the issue of educational effectiveness, educational research must adopt the paradigm of institutional complexity that allows it to examine causal relationships realistically and in a manner compatible with the nature of schools and their purposes [48] [49]. Authors such as Radford [48] and Galán, Ruiz-Corbella & Sánchez Mellado [49] argue that educational research must be approached from the paradigm of complexity and chaos. The theory of complexity and chaos postulates that a system is complex when its operation focuses on multiple relationships and with multiple components, interacting simultaneously [49]. Complex systems always operate in a dichotomy between order and disorder. Order is the desired functioning of the system and disorder always constitutes an opportunity for improvement and innovation to develop the system when it decomposes, deviates or is aspired to higher levels of effectiveness. Complex systems show effects and products that are not always equitable to the "cause" because their relationships are not always linear, orderly and logical. This has been evident in the study of the organization of groups of certain species of animals, such as ants, where teamwork is the axis of functioning. For example, in the presence of danger, a group member may emerge as a leader and undergo an internal reorganization of the system and group. These multidimensional relationships can only be understood if they are studied in the context of the organization. Improving these interactions and relationships can only be achieved in the context of organizational structures. Another element that affects complex systems is the time variable. This implies that certain operations and certain tasks must occur at certain dates for the effective and desired performance that results in the stability of the system.

In the pursuit of educational efficiency, educational research has to approach schools as complex systems. It needs to be understood that the product or outcomes of complex systems such as school do not come from a single set of causes but from the sum of many multidimensional relationships that are not always linear. This implies a redefinition of the research approach of the school effectiveness that recognizes the functioning and the complexity of the reality of the schools. This does not rule out linear research on causes and effects that explain the results of standardized tests, but provides the space for those who wish to investigate education as an instrument for student development, they can do so [37]. In the 21st century, educational research needs a paradigm that organizes it to address the complexity of education [20] [48] [49]. Being able to understand the functioning and effect or results of the multiple relationships that occur in schools in the context of their organizational, administrative and policy structures should facilitate improving the efficiency of schools and education systems. This should facilitate improving the practice of education and educational policies in the context of the complexity of education systems.

Strategy 8: Link research to educational policies
Given its political nature in the field of education, its practice needs to move to a science-based one that helps to eradicate practices and educational models centered on partisan ideologies and politics. Linking research to the formulation of educational policies is an alternative to scientifically improving the field of education. It would be very necessary, in addition, its relation with teacher training [40] [50]. Educational policies for social service must understand the importance of educational research at all levels, not only the university, and encourage its practice in teachers and professionals in education in practice.

Strategy 9: Increase ICT in educational research
The emergence of ICT in the last decades offers new possibilities for educational research, especially in terms of data management. ICT in education research presents great opportunities to enter the complexity of education and generate more accurate and reliable data. Its potential for improving research effectiveness is a developing topic. It may seem that the traditional aspiration of the social sciences, including the educational sciences, is achieved, that of approaching the methods of the experimental sciences with the possibility of quantifying high volumes of information of all kinds and variables. Although we can find several studies that analyze the possible drawbacks of their real effectiveness in educational research [36] [51] [52] [53] [54] [55] [56] [57], we consider that the main problem that we face is that these new possibilities offered by ICTs to develop empirical methodologies, with great capacity in data management, are conditioning the nature and the objectives of serving authentic educational needs [36] [38]. The current characteristics of science and scientific diffusion lead many educational researchers, in order to promote the publication of their works, to focus on the method above the ends and the goals of a real educational utility, which provides an authentic social and pedagogical benefit. Irrefutable research in their methods and data management, which with ICT are presented with an extraordinary rigor and quality, but of dubious contribution to educational knowledge.

There is no doubt that the use of ICT in educational research can offer extraordinary possibilities, but without losing sight of the fact that they must serve the objectives pursued. ICT should be a useful tool for the process but not become the end of it. Their presence, moreover, in educational research should not only be produced from a technical perspective. When we find ourselves in a digital society dominated by them, it is imperative that educational researchers confront as a challenge the analysis and critical study of this reality and its true impact on education [55] [56] [58]. Encouraging the integration of ICTs, which will enable them to become more knowledgeable, will contribute decisively to this. Multiple models of educational technology, from a molecular or global dimension, can help a growing presence of studies that consider them from different perspectives multidisciplinary. The integration of ICT in educational research should be an exercise to be increasingly considered in the planning of research processes in the field of education.

## 3. Conclusions

The foundations of contemporary educational research need to be revised and rethought. We have offered different proposals that lead us to new research frameworks in the field of educational sciences, parallel to the new concepts present in the philosophy of science and the need to improve the knowledge processes for a better practical application of them.

In brief, the nine strategies of scientific effectiveness are: a) define and demarcate the concept of educational research, b) define the education construct as a research phenomenon, c) conduct research that is useful for education, d) increase the validity of educational research, e) Increase the generalization of educational research, f) careful selection of the research design to be used, g) explore new models for the study of causal relationships and educational effectiveness, h) link research to educational policies and, i) Increase ICT in educational research.

We understand that despite all of them are challenging, we have great opportunities for develop and achieve them. Finally, although the road map that emerges from the discussion that we present does not even show the paved roads that are aspirated to, the sidewalks that are delineated allow walking forward with a better sense of direction.

**References**

[1] A. E Blumberg & H. Feigl, *The Journal of Philosophy*, 28(11) (1931) 281-296
[2] F. Stadler, *The Vienna Circle: Studies in the Origins, Development, and Influence of Logical Empiricism*, Springer, Vienna, 2001
[3] T. Uebel, *Perspectives on Science* 21(1) (2013) 58-99
[4] P. Feyerabend, *Against Method*, Verso, London, 1975
[5] M. T. Gargiulo, *Revista Colombiana de Filosofía*, 65(160) (2016) 95-120
[6] J. D. Cifuentes, *Escritos*, 17(39) (2010) 494-518
[7] L. Garagalza, *Revista Portuguesa de Filosofía*, 43 (1987) 177-190
[8] R. S. Cohen, *Synthese*, 18(2) (1968) 132-170
[9] M. J. Marr, *Behavior and Philosophy*, 31 (2003) 181-192
[10] J. Blackmore, *The British Journal for the Philosophy of Science* 36(3) (1985) 299-305
[11] K. Popper, Realism and the Aim of Science: From the Postscript to the Logic of Scientific Discovery. Routledge, London and New York, 1983
[12] D. Rowbottom, *Popper's Critical Rationalism: a Philosophical Investigation*, Routledge, New York, 2011
[13] J. Gómez Galán, *A New Educational Research to Society, Citizens and Human Development*. J. Gómez Galán, E. López Meneses & A. H. Martín (eds.), UMET Press, Cupey, 2015, 7-12
[14] J. Gómez Galán, *International Journal of Educational Research and Innovation*, 6 (2016) 124-145
[15] O. Ponce, N. Pagán & J. Gómez Galán, *Filosofía de la Investigación Educativa en una Era Global: Retos y Oportunidades de Efectividad Científica*, Publicaciones Puertorriqueñas, San Juan, 2017
[16] R. Pring, *Philosophy of Educational Research* (2[nd] ed), Continuum, London, 2000
[17] A. Ellis, *Research on Educational Innovation*. Routledge. London and New York, 2014
[18] D. C. Phillips, *Journal of Philosophy of Education*, 39(4) (2005) 577-597.
[19] C. Thompson, *Studies in Philosophy and Education*, 31(3) (2012) 239-250.
[20] O. A. Ponce, *Investigación Educativa*, Publicaciones Puertorriqueñas Inc San Juan, Puerto Rico, 2016
[21] O. A. Ponce, & N. Pagán-Maldonado, *International Journal of Educational Excellence,* 1(1) (2015) 111-135.
[22] D. C. Phillips, *A Quixotic Quest? Philosophical Issues in Assessing the Quality of Educational Research*, P. Walters, A. Lareau & S. Ranis (ed.), Routledge, New York & Londres, 2009
[23] O. A. Ponce, *Investigación de Métodos Mixtos en Educación*, Publicaciones Puertorriqueñas Inc., Hato Rey, Puerto Rico, 2014
[24] O. A. Ponce, *Investigación Cualitativa en Educación: Teoría, Prácticas y Debates*, Publicaciones Puertorriqueñas Inc., San Juan, Puerto Rico, 2014
[25] O. A. Ponce, & N. Pagán-Maldonado, *Investigación Educativa: Retos y Oportunidades,* J. Gómez Galán, E. López Meneses & L. Molina (eds.), *Research Foundations of the Social Sciences,* UMET Press Cupey, 2016, 110-121
[26] J. M. Touriñán, *Revista de Investigación en Educación*, 12(1) (2014) 6-31.
[27] J. Gómez Galán, (ed.), *Educational Research in Higher Education: Methods and Experiences*, River Publishers, Aalborg, 2016
[28] M. Koichiro, *Educational Studies in Japan: International Yearbook*, 7(3) (2013) 37-49
[29] P. Smeyers, *Journal of Philosophy of Education*, 47(2) (2013) 311-331.
[30] A. Lee, *The Australian Educational Researcher*. 37(4) (2010) 63-78.
[31] D. Hargreaves, *Teaching as Research-Based Profession: Possibilities and Prospects*, M. Hammersley,

[32] J. Peñalva, *Journal of Philosophy of Education*, 48(3) (2014) 400-415.
[33] J. Scheneider, *Phi Delta Kappan,* 96(1) (2014) 30-35
[34] L. Lysenko, P. Abrami, R. Bernand, C. Degenais, & M. Janosz, *Canadian Journal of Education*, 37(2) (2014) 1-26.
[35] S. H. Ranis, *Blending Quality and Utility: Lessons Learned from the Educational Research Debates*, P. B. Walter, A. Lareau, & S. H. Ranis, Taylor & Francis Group, New York and London, 2009
[36] J. Gómez Galán, *Social Sciences Research in the ICT Age*, J. Gómez Galán, E. López Meneses & L. Molina (eds.), UMET Press, Cupey, 2015, 6-10
[37] G. Biesta, *European Educational Research Journal*, 14(1) (2015) 11-22
[38] J. Gómez Galán, *The Importance of Real Pedagogical Training in New Teaching Strategies for Quality Education*, J. Gómez Galán, E. López Meneses & L. Molina, (eds.), Instructional Strategies in Teacher Training UMET Press, Cupey, 2015, 9-14
[39] S. C. Marley, & J. R. Levin, *Educational Psychology Review*, 23(2) (2011) 197-206.
[40] C. Winch, A. Oancea, & J. Orchard, *Oxford Review of Education*, 41(2) (2015) 202-216.
[41] M. A. Vinovskis, A History of Efforts to Improve the Quality of Federal Education Research: From Gardner's Task Force to the Institute of Education Science, P. Walter, A. Lareau & S. H. Ranis (eds), Taylor & Francis Group, New York and London, 2009
[42] C. Clark, *Journal of Philosophy of Education*, 45(1) (2011) 37-57.
[43] S. M. Ravitch, *Perspective of Urban Education,* 11(1) (2014) 1-10.
[44] A. Mejías, *Studies in Philosophy and Education*, 27(2-3) (2008) 161-171.
[45] J. Schofield, *Increasing Generalizability of Qualitative Research*, M. Hammersley (ed), Sage Publications, Los Angeles, 2007
[46] D. Berliner, *Educational Researcher*. 31(8) (2002) 18-20
[47] F. Gil-Cantero, & D. Reyero, *Revista Española de Pedagogía,* 258 (2014) 263-280
[48] M. Radford, *British Educational Research Journal,* 32(2) (2006) 177–190
[49] A. Galán, M. Ruíz-Corbella, & J. C. Sánchez Mellado, *Revista Española de Pedagogía*, 258 (2014) 281-298
[50] N. Fancourt, *Journal of Beliefs & Values*. 37(1) (2016) 123-124
[51] J. Gómez Galán, *Corrientes de Investigación en Tecnología Educativa*, J. Gómez Galán & G. Lacerda (eds), Liber Livro Editora/Universidade de Brasília, Brasília, 2012, 105-159
[52] N. Selwyn, *Learning, Media and Technology*, 40(1) (2015) 64-82.
[53] J. Gómez Galán, *Paradigmas de Investigación Científica en Tecnología Educativa: Evolución Histórica y Tendencias Actuales*, F. J. Durán (ed.), ACCI, Madrid, 2014, 195-224
[54] D. Lundie, *Educational Philosophy and Theory*, 1 (2015) 1-14.
[55] J. Gómez Galán, *Educar en Nuevas Tecnologías y Medios de Comunicación*, Fondo Educación CRE, Sevilla-Badajoz, 2003
[56] J. Gómez Galán, *Impacto de la Sociedad Tecno-Mediática en la Educación: Estrategias Transformadoras en la Formación del Profesorado*, J. Cabero *et al.* (eds.), Minerva Universidad, Seville, 2011, 17-35
[57] S. Manca, L. Caviglione, & J. E. Raffaghelli, *Journal of e-Learning and Knowledge Society*, 12(2) (2016) 27-39.